\documentclass[a4paper,aps,prd,10pt,preprintnumbers,showkeys,twocolumn,superscriptaddress,nofootinbib,amsmath,amssymb]{revtex4-1}
\usepackage{cmap,graphicx,orcidlink}
\usepackage[utf8]{inputenc}
\usepackage[T1]{fontenc}

\def\Order#1{{\cal O}\left(#1\right)}

\begin{document}

\title{Primary hairs may create echoes}

\author{R. A. Konoplya \orcidlink{0000-0003-1343-9584}}
\email{roman.konoplya@gmail.com}
\affiliation{Research Centre for Theoretical Physics and Astrophysics, Institute of Physics, Silesian University in Opava, Bezručovo náměstí 13, CZ-74601 Opava, Czech Republic}

\author{A. Zhidenko \orcidlink{0000-0001-6838-3309}}
\email{olexandr.zhydenko@ufabc.edu.br}
\affiliation{Research Centre for Theoretical Physics and Astrophysics, Institute of Physics, Silesian University in Opava, Bezručovo náměstí 13, CZ-74601 Opava, Czech Republic}
\affiliation{Centro de Matemática, Computação e Cognição (CMCC), Universidade Federal do ABC (UFABC), \\ Rua Abolição, CEP: 09210-180, Santo André, SP, Brazil}

\begin{abstract}
In most scenarios studied so far, the appearance of echoes in the ringdown signal requires modifications external to the black hole itself, such as the presence of matter in the near-horizon region, quantum field clouds, or exotic compact objects like wormholes that effectively introduce additional peaks in the effective potential. In this work we show that echoes can naturally arise in a different setting: black holes endowed with primary Proca-Gauss-Bonnet hair. We demonstrate that the primary hair modifies the effective potential in such a way that a second peak is formed, giving rise to late-time echoes without invoking any external environment or exotic horizon-scale physics. Using both the higher-order WKB method with Padé resummation and time-domain integration, we compute the quasinormal spectrum for scalar and Dirac test fields and show the appearance of these echoes. Our results highlight a novel mechanism by which primary hairs alone can leave observable imprints on the ringdown signal of black holes in modified gravity.
\end{abstract}


\maketitle

\section{Introduction}

The study of linear perturbations around black holes provides a powerful probe of their stability, structure, and potential deviations from general relativity. In particular, the spectrum of quasinormal modes --- complex frequencies characterizing the damped oscillations of perturbed black holes --- serves as a unique fingerprint of the background spacetime. With the increasing precision of gravitational wave observations, quasinormal modes have become a key observable for testing modifications of Einstein gravity in the strong-field regime~\cite{Berti:2009kk,Konoplya:2011qq,Cardoso:2016ryw,Bolokhov:2025uxz}.

A number of modified gravity theories predict the existence of black holes endowed with additional degrees of freedom, such as scalar or vector fields, which can influence their quasinormal spectrum. Of particular interest are those theories that allow for analytical black hole solutions incorporating non-trivial \textit{primary hair} --- global charges or integration constants not associated with standard conserved quantities like mass or electric charge. These models are especially suitable for the analysis of the quasinormal modes, as they provide explicit backgrounds for studying how new physics modifies the ringdown signals.

In~\cite{Charmousis:2025jpx}, a novel class of spherically symmetric black holes was obtained in a theory combining scalar-tensor and vector-tensor interactions of Gauss-Bonnet type. The action of the model is given by
\begin{equation}\label{action}
S = \int d^4x \sqrt{-g} \left( R - \alpha\, \mathcal{L}^{\text{VT}}_G - \beta\, \mathcal{L}^{\text{ST}}_G \right),
\end{equation}
where
\begin{equation}
\mathcal{L}_\mathcal{G}^{\rm VT} = 4 G^{\mu \nu}W_\mu W_\nu + 8 (W_\nu W^\nu) \nabla_\mu W^\mu + 6(W_\nu W^\nu)^2
\end{equation}
represents the vector-tensor theory coupling with the vector field \(W_\mu\) and
\begin{equation}
\mathcal{L}_\mathcal{G}^{\rm ST} = \phi \mathcal{G} - 4 G^{\mu \nu} \nabla_\mu \phi \nabla_\nu \phi - 4 \Box \phi (\partial \phi)^2 - 2 (\partial \phi)^4.
\end{equation}
is the scalar-tensor counterparts with the scalar field \(\phi\) \cite{Fernandes:2021dsb}. Here
\begin{equation}
\mathcal{G} = R^2 - 4 R_{\mu \nu} R^{\mu \nu} + R_{\mu \nu \alpha \beta} R^{\mu \nu \alpha \beta}
\end{equation}
is the Gauss-Bonnet term and the Einstein tensor is defined as follows:
\begin{equation}
G_{\mu \nu} = R_{\mu \nu} - \frac{1}{2}g_{\mu\nu}R.
\end{equation}

What makes these solutions particularly appealing is their regularity and asymptotic flatness, along with the presence of a vector Proca field that sources the geometry independently of the mass \cite{Kase:2018owh}. Depending on the relation between the couplings \( \alpha \) and \( \beta \), the spacetime metric admits either a rational form (when \( \beta = -\alpha \)) or a square-root structure (for generic values), with both families reducing to Schwarzschild in the appropriate limit. These properties offer a clean setting for isolating the effects of Proca hair on black hole dynamics.

The particle motion, shadows, and grey-body factors for these black holes were recently studied in \cite{Lutfuoglu:2025ldc}, but an analogous analysis of their quasinormal spectrum has not yet been carried out. In this paper, we demonstrate that the quasinormal ringing of such black holes possesses a remarkable property: within certain ranges of parameters it can produce echoes without the need for any environmental effects in the far zone or specific quantum corrections near the event horizon.

Echoes arise from secondary scattering whenever the effective potential develops more than one peak. Typically, an additional peak is introduced artificially, for instance, to model the influence of the astrophysical environment, the presence of a cloud of quantum fields near the horizon, or exotic compact objects such as wormholes that mimic black-hole behavior by duplicating the potential barrier \cite{Cardoso:2016rao,Cardoso:2016oxy,Abedi:2016hgu,Mark:2017dnq,Bueno:2017hyj,Konoplya:2018yrp,Cardoso:2019apo,Bronnikov:2019sbx,Dong:2020odp,Churilova:2021tgn,Cheung:2021bol,Konoplya:2024wds}. Another way to obtain secondary scattering is through the AdS boundary, which has been studied in particular for scalar-tensor theories of gravity \cite{Vlachos:2021weq,Chatzifotis:2021pak}. As a rule, black hole solutions in Einstein gravity or its common extensions do not give rise to echoes in asymptotically flat spacetime. Here, however, we show that \emph{primary hairs alone can induce a nonmonotonic behavior of the metric function, which in turn generates strong echoes.}

The paper is organized as follows. In Secs.~\ref{sec:background} and~\ref{sec:equations} we present the background geometry and the wave equations for scalar and Dirac perturbations. Section~\ref{sec:numerical} provides a brief overview of the numerical methods employed to compute quasinormal modes and to study the evolution of perturbations, namely the higher-order WKB method with Padé approximants and the time-domain integration. The results on echoes and quasinormal modes are discussed in Secs.~\ref{sec:echoes} and~\ref{sec:qnms} respectively. Finally, Sec.~\ref{sec:conclusions} contains our conclusions, where we summarize the main findings and outline some open questions.

\section{Background geometry}\label{sec:background}

We investigate the quasinormal modes of two static, spherically symmetric black hole solutions derived in~\cite{Charmousis:2025jpx}, which arise from the action~\eqref{action} incorporating both vector-tensor and scalar-tensor Gauss-Bonnet-type interactions. The line element common to both configurations is expressed as
\begin{equation}
ds^2 = -f(r)\, dt^2 + \frac{dr^2}{f(r)} + r^2 (d\theta^2 + \sin^2 \theta\, d\phi^2),
\end{equation}
where the functional form of \( f(r) \) depends on the values of the coupling constants \( \alpha \) and \( \beta \).

In the generic case \( \alpha \neq \beta \), the metric function takes the form
\begin{align}\label{metricfunc}
f(r) &= 1 - \frac{2 \alpha (M - Q)}{r (\alpha + \beta)} + \frac{r^2}{2 (\alpha + \beta)} \\\nonumber
& - \frac{r^2}{2 (\alpha + \beta)}
\sqrt{1 + \frac{8 \alpha Q}{r^3} + \frac{8 \beta M}{r^3} - \frac{16 \alpha \beta (M - Q)^2}{r^6} },
\end{align}
with \( M \) denoting the ADM mass and \( Q \) representing an independent integration constant associated with the Proca field. Crucially, \( Q \) is not constrained by \( M \), indicating the presence of \emph{primary hair} --- an intrinsic attribute of the black hole that cannot be reduced to standard conserved charges. When \( Q = M \), the geometry coincides with the black hole solution found in the scalar-tensor sector of four-dimensional Einstein-Gauss-Bonnet gravity~\cite{Lu:2020iav,Kobayashi:2020wqy,Fernandes:2020nbq}.

In the limit \( \beta \to -\alpha \), the expression~\eqref{metricfunc} simplifies considerably, yielding a rational form:
\begin{align}
\lim_{\beta\to-\alpha}f(r) &= \frac{r^3}{r^3 - 4 \alpha (M - Q)} \Bigg( 1 - \frac{2 M}{r}
 \\\nonumber
&\qquad + \frac{4 \alpha (M - Q)^2}{r^4}
- \frac{4 \alpha (M - Q)}{r^3} \Bigg).
\end{align}
This solution remains regular provided the denominator does not vanish outside the event horizon. In both cases, the Schwarzschild geometry is recovered in the limit \( \alpha = \beta = 0 \).

The position of the event horizon, \(r_h>0\), if present, is determined by solving the equation \( f(r_h) = 0 \). The number and existence of horizons depend sensitively on the parameters \( \alpha \), \( \beta \), \( M \), and \( Q \). At spatial infinity, both spacetimes asymptote to Schwarzschild,
\[
f(r) = 1 - \frac{2 M}{r} + \Order{r^{-2}},
\]
confirming that \( M \) corresponds to the ADM mass~\cite{Arnowitt:1960es}.

The full parameter space compatible with black hole solutions is explored in~\cite{Charmousis:2025jpx}. It was shown there that large positive values of the couplings \( \alpha \) and \( \beta \) may inhibit horizon formation when \( \alpha + \beta \neq 0 \), and that similar obstructions occur for sufficiently large values of \( Q \) in the \( \alpha + \beta = 0 \) case. In such regimes, the spacetime instead describes a naked singularity or a horizonless compact object, depending on the parameter configuration.

\section{Perturbation equations}\label{sec:equations}

To study the quasinormal spectrum of the black hole solutions described in the previous section, we consider test fields propagating on the fixed background spacetime. Specifically, we analyze massless scalar, and Dirac fields. In both cases, the field equations reduce to a Schrödinger-like master equation of the form
\begin{equation}\label{master_eq}
\frac{d^2 \Psi}{dr_*^2} + \left[ \omega^2 - V(r) \right] \Psi = 0,
\end{equation}
where \( \omega \) is the complex quasinormal frequency and \( r_* \) is the tortoise coordinate, defined via
\begin{equation}
\frac{dr_*}{dr} = \frac{1}{f(r)}.
\end{equation}

The effective potential for a massless scalar field is
\begin{equation}
V(r) = f(r) \left( \frac{\ell(\ell+1)}{r^2} + \frac{f'(r)}{r} \right),
\end{equation}
where \(\ell=0,1,2,3,\ldots\) is the multipole number.

\begin{figure*}
\resizebox{\linewidth}{!}{\includegraphics{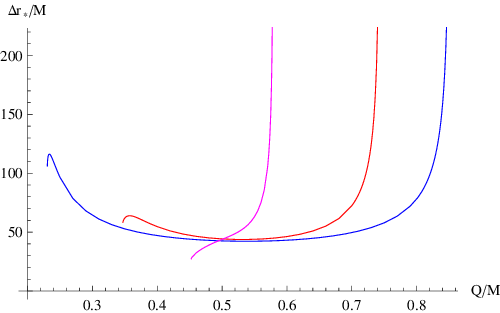}\includegraphics{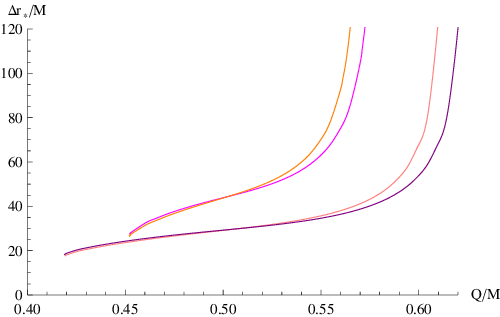}}
\caption{Distance between the main and secondary peak of the effective potential \(s=\ell=0\) as a function of $Q$. Left panel for fixed \(\beta=1.05M^2\) (from left to right): \(\alpha=-M^2\) (magenta), \(\alpha=-0.2M^2\) (red), \(\alpha=-0.1M^2\) (blue). Right panel (from left to right): \(\alpha=-1.1M^2\) and \(\beta=1.05M^2\) (orange), \(\alpha=-M^2\) and \(\beta=1.05M^2\) (magenta), \(\alpha=-1.1M^2\) and \(\beta=1.1M^2\) (pink), \(\alpha=-M^2\) and \(\beta=1.1M^2\) (purple).}\label{fig:peaks}
\end{figure*}

In the case of a massless Dirac field, the decoupled radial equations for the two-component spinor yield two effective potentials given by \cite{Cho:2003qe}
\begin{equation}
V(r) = f(r)\frac{\kappa^2}{r^2} \pm f(r) \frac{d}{dr} \left( \frac{\kappa\sqrt{f(r)}}{r} \right),
\end{equation}
where \( \kappa\equiv\ell + 1 =1,2,3,\ldots\).

Since these two potentials are isospectral, it is sufficient to use only one of the them for calculations of the quasinormal modes. We use here only the ``\(-\)'' potential, because the WKB method is usually more accurate for such choice. The effective potentials for all spins exhibit an additional peak within a certain range of black hole parameters (see Fig.~\ref{fig:peaks}), which will be discussed in detail below.

The boundary conditions for quasinormal modes in asymptotically flat spacetimes are purely ingoing waves at the event horizon ($r=r_h$) and purely outgoing waves at spatial infinity:
\begin{equation}
\begin{array}{rlcl}
\Psi(r_*) &\propto e^{-i\omega r_*}, &\quad& r_* \to -\infty \quad (r \to r_h), \\
\Psi(r_*) &\propto e^{+i\omega r_*}, &\quad& r_* \to +\infty \quad (r \to \infty).
\end{array}
\end{equation}

\section{Numerical methods}\label{sec:numerical}

Here we discuss two complementary numerical approaches for studying the quasinormal spectrum in both the frequency and time domains: the higher-order WKB method with Padé approximants, and the time-domain integration technique.

\subsection{Time-domain integration}

In order to obtain the time-domain profile we employ the Gundlach-Price-Pullin method \cite{Gundlach:1993tp}. We solve the wavelike equation (\ref{master_eq}) using the null coordinates, \(u = t - r_*\) and \(v = t + r_*\),
\[
4\frac{\partial^2 \Psi}{\partial u\partial v} + V(r_*)\Psi = 0.
\]
The wave equation is discretized on a numerical grid, and the following integration scheme is used
\begin{eqnarray}\label{Discretization}
\Psi\left(N\right)&=&\Psi\left(W\right)+\Psi\left(E\right)-\Psi\left(S\right)
\\&&\nonumber
-h^2V\left(S\right)\frac{\Psi\left(W\right)+\Psi\left(E\right)}{8}+\Order{h^4}.
\end{eqnarray}
Here, \(N\equiv\left(u+h,v+h\right)\), \(W\equiv\left(u+h,v\right)\), \(E\equiv\left(u,v+h\right)\), and \(S\equiv\left(u,v\right)\) represent neighboring points on the computational grid. The initial conditions are specified on the null surfaces, \(u=u_0\) and \(v=v_0\). We adopt Gaussian wave packets on both surfaces, centered at \((u,v)=(u_0,v_0)\), and localized near the maximum of the effective potential. With this choice, the resulting evolution quickly settles into the characteristic ringdown profile after a very short period of initial outburst \cite{Konoplya:2011qq}.

This scheme captures the full time-domain evolution, including the initial burst, the quasinormal ringing phase, and the late-time tails. Once the time-domain profiles of the perturbations are obtained, the Prony method is employed to extract the quasinormal frequencies. The method fits the time-domain signal \(\Psi(t)\) to a sum of exponents:
\[
\Psi(t) = \sum_{k=1}^N A_k e^{-i\omega_k t},
\]
where \(\omega_k\) represents the complex frequency. By solving the least squares problem for the coefficients \(A_k\) and frequencies \(\omega_k\), the dominant modes can be determined with a good accuracy.

\begin{figure*}
\resizebox{\linewidth}{!}{\includegraphics{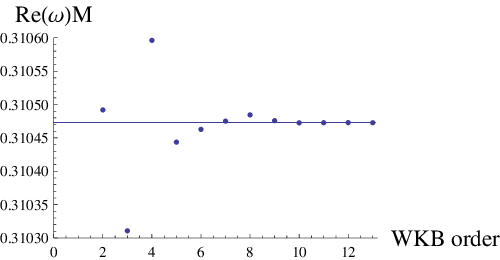}\includegraphics{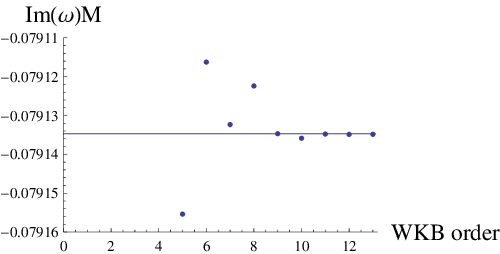}\includegraphics{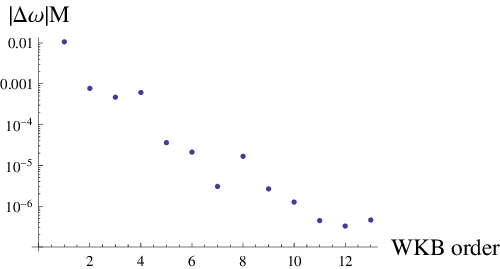}}
\caption{Dominant quasinormal frequency of the scalar field (\(\ell=1\)) for \(\alpha=-1.1M^2\), \(\beta=1.03M^2\), \(Q=0.56M\) calculated with the help of the WKB formula with the Padé approximants vs the Prony fit of the time-domain profile at late times of the ringdown (solid line), resulting to an accurate value of the fundamental frequency, \(\omega M\approx0.3104732-0.0791347i\). The right panel shows absolute deviation from the time-domain result on semilogarithmic scale.}\label{fig:WKBconvergence}
\end{figure*}

\subsection{WKB Method with Padé resummation}

To compute the quasinormal mode frequencies, we employ the Wentzel-Kramers-Brillouin (WKB) method, which has been widely used for black hole perturbations~\cite{Schutz:1985km,Iyer:1986np,Konoplya:2003ii}. In this approach, one matches approximate solutions near the peak of the potential barrier to the asymptotic regions.

This procedure is particularly effective when the effective potential has the form of a single-peaked barrier, which is the typical situation for perturbations of black holes. Even if the potential develops a small negative gap close to the horizon, the WKB approach often remains a reliable tool for estimating the dominant quasinormal frequencies.

At the lowest order, the WKB formula reduces to the eikonal approximation, which becomes exact in the limit of large multipole number $\ell \to \infty$. More generally, the WKB expansion provides an analytic expression for the quasinormal frequencies as a series in powers of the inverse eikonal parameter
\begin{equation}
\frac{i (\omega^2 - V_0)}{\sqrt{-2 V_0''}} - \sum_{j=2}^{k} \Lambda_j = n + \frac{1}{2}, \quad n = 0,1,2,\dots
\end{equation}
where with \(n=0,1,2,3,\ldots\) is the overtone number, \( V_0 \) is the maximum of the effective potential \( V(r) \), and \( V_0'' \) is the second derivative with respect to \( r_* \) evaluated at the peak. The \( \Lambda_j \) are higher-order correction terms involving derivatives of \( V(r) \) up to order \( 2j \). Explicit expressions for \( \Lambda_j \) have been computed up to \( k = 13 \) in~\cite{Matyjasek:2017psv}.

In order to improve the convergence of the WKB approximation ---especially for low multipole numbers \( \ell \) we adopt Padé resummation~\cite{Konoplya:2019hlu}, which reorganizes the WKB series into a rational function with respect to the bookkeeping parameter \(\epsilon\), which is introduced assuming that \(\Lambda_j\propto\epsilon^j\),
\begin{equation}
P_{\tilde{n}/\tilde{m}}=\frac{Q_0+Q_1\epsilon+\ldots+Q_{\tilde{n}}\epsilon^{\tilde{n}}}{R_0+R_1\epsilon+\ldots+R_{\tilde{m}}\epsilon^{\tilde{m}}},
\end{equation}
where \(\tilde{n}+\tilde{m}=k\).

The best accuracy within a given WKB order \(k\) is usually achieved when \(\tilde{n} \approx \tilde{m} \approx k/2\) \cite{Matyjasek:2017psv,Konoplya:2019hlu}. To test convergence, we adopt the convention that for each order \(k\) we use Padé approximants with \(\tilde{m} = \tilde{n}\) when \(k\) is even, and \(\tilde{m} = \tilde{n} + 1\) when \(k\) is odd. For instance, for \(k=12\) we employ \(P_{6/6}\), while for \(k=13\) we use \(P_{6/7}\). Figure~\ref{fig:WKBconvergence} illustrates the convergence of the fundamental frequency as a function of the order $k$ for the test scalar field with \(\ell=1\).

We observe that increasing the WKB order does not necessarily guarantee systematic improvement, but in general higher orders tend to provide better approximations. For \(\ell=1\), the 13th-order WKB formula with the Padé approximant \(P_{6/7}\) achieves an accuracy of about 5-7 decimal places. For \(\ell=0\), the convergence is slower, but still sufficient to determine 3 decimal places of the fundamental mode. For the Dirac field \(\kappa=1\) the Padé approximant \(P_{6/7}\) of the WKB value allows for calculation of 3-4 decimal places. This level of precision is adequate for detecting the effects induced by the Proca-Gauss-Bonnet hair studied in the present work.

\begin{figure*}
\resizebox{\linewidth}{!}{\includegraphics{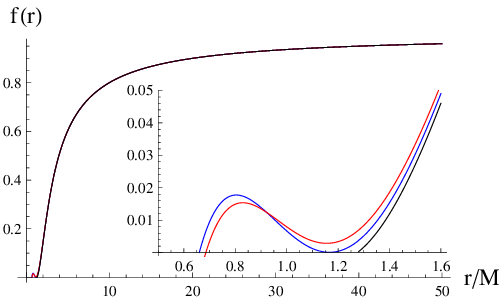}\includegraphics{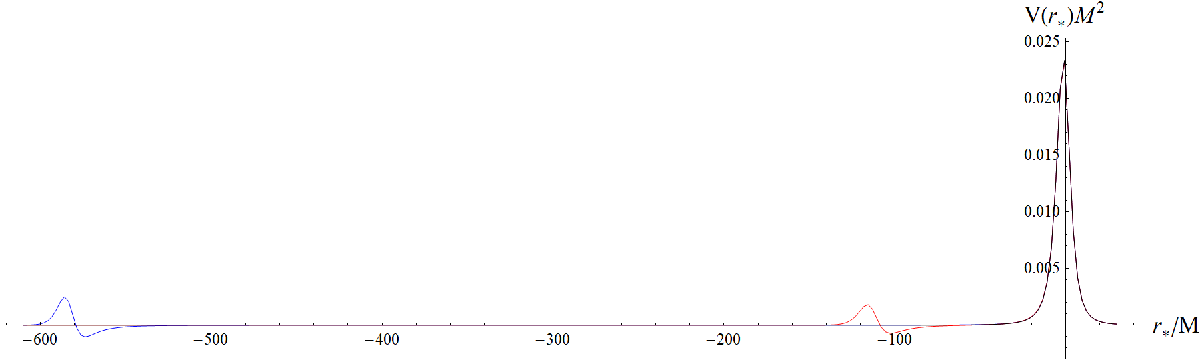}}
\resizebox{\linewidth}{!}{\includegraphics{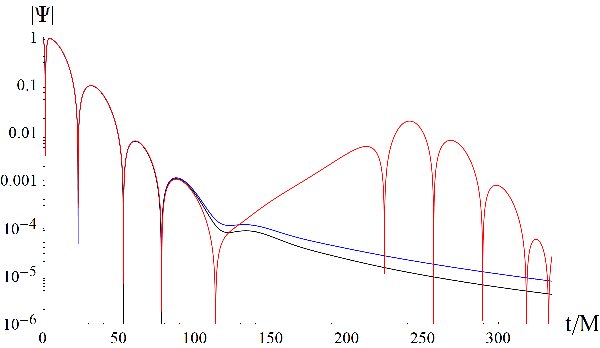}\includegraphics{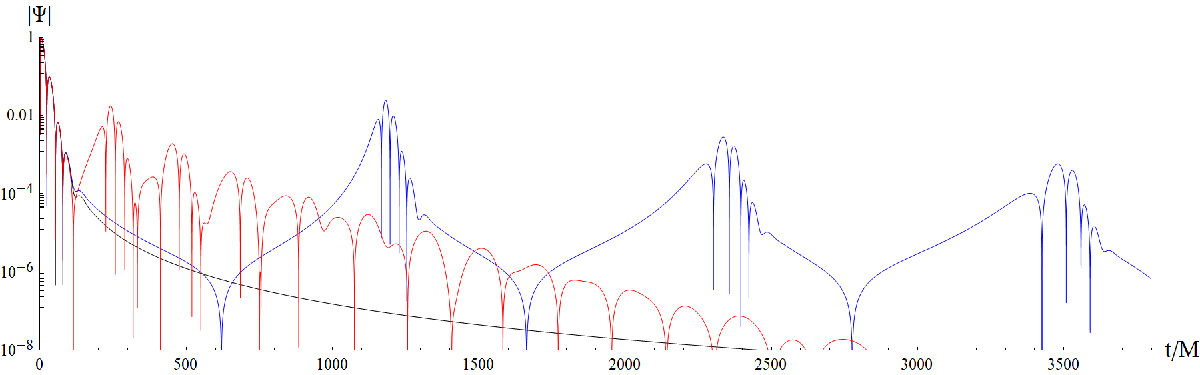}}
\resizebox{\linewidth}{!}{\includegraphics{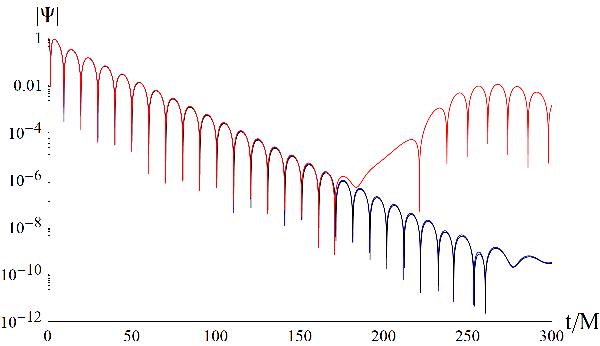}\includegraphics{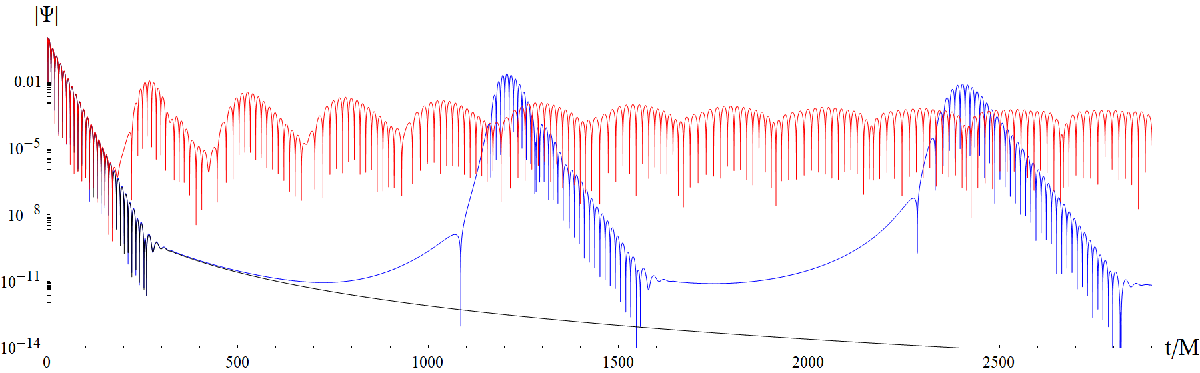}}
\caption{Metric and time-domain profiles of the test scalar field for the black holes with
\(\alpha=-1.1M^2\) and \(\beta=1.03M^2\): \(Q=0.54M\) (red), \(Q=0.55M\) (blue), and \(Q=0.56M\) (black). The \emph{upper-left panel} shows the metric function starting from the horizon. For the first two black-hole configurations, the nonmonotonic behavior of the metric function produces an additional (smaller) peak in the effective potential. The \emph{upper-right panel} illustrates this feature for \(s=\ell=0\); for other fields and multipole numbers, the picture is qualitatively similar. The \emph{middle panels} display \(\ell=0\) perturbation profiles: at early times, the perturbations exhibit a ringdown governed by the main peak, which is nearly identical for the black holes considered. For the second configuration, echoes appear only at very late times, consistent with the large separation between the two peaks. In the first configuration, the shorter separation leads to more frequent echoes, followed by the final relaxation governed by a dominant quasinormal frequency different from that of the ringdown phase. The \emph{bottom panels} present \(\ell=1\) perturbation profiles: the echoes have nearly the same time interval, since the peak separation is similar; however, because the main peak is higher than in the \(\ell=0\) case, the ringdown stage lasts longer and the final relaxation occurs much later. For the third configuration, there is no additional peak, and no echoes are observed: after the ringdown, the signal transitions into the asymptotic power-law tail.}\label{fig:echoappearing}
\end{figure*}

\section{Echoes due to the Proca-Gauss-Bonnet hair}\label{sec:echoes}

The family of black-hole solutions in Proca-Gauss-Bonnet gravity exhibits a parametric region in which the metric function \(f(r)\) becomes nonmonotonic. For certain values of the coupling parameters \(\alpha\) and \(\beta\), and for black-hole hair parameter \(Q\) (in units of mass, \(M=1\)), the function \(f(r)\) develops a local maximum followed by a local minimum near the horizon. As \(Q\) increases, the value of \(f(r)\) at the local minimum decreases, and at a critical value of \(Q\) it reaches zero. At this point, the event-horizon radius undergoes a discontinuous change as a function of \(Q\). For \(Q\) larger than the critical value, the minimum shifts inside the new horizon.

This nonmonotonic behavior of the metric function generates an additional peak in the effective potential of the black-hole perturbations (see Fig.~\ref{fig:echoappearing}). As the local minimum of \(f(r)\) approaches zero, the corresponding secondary peak of the effective potential in the tortoise coordinate moves further from the main peak toward the horizon, thereby increasing the time interval between successive echoes. Beyond the critical value of \(Q\), the additional peak disappears, and no echoes are produced. Conversely, when the hair parameter \(Q\) decreases, the nonmonotonic region of the metric function shifts toward the horizon and eventually vanishes. In this regime, the amplitude of the secondary peak diminishes and finally disappears, so that echoes are again absent for sufficiently small values of \(Q\).

Such a parametric region exists for sufficiently large values of the coupling \(\beta > M^2\) or for small black holes. The other coupling parameter \(\alpha\) must be negative so that both the scalar–tensor and vector–tensor couplings contribute to producing the secondary peak. When \(\alpha\) is only slightly negative, \(\alpha \gtrsim -0.03M^2\), the corresponding range of \(Q\) includes both positive and negative values. However, as \(\alpha\) decreases further, the interval of the values of the hair parameter \(Q\) that allow for nonmonotonic behavior becomes narrower. In Fig.~\ref{fig:peaks} we present the distance between the two peaks for various fixed values of \(\alpha\) and \(\beta\). The upper (right) bound of the region corresponds to the point where the local minimum of the metric function vanishes, which leads to a discontinuity in the event-horizon coordinate. At this value of \(Q\), the distance between the two peaks diverges, producing an infinite asymptote. The lower (left) bound of the region corresponds to the disappearance of the nonmonotonic behavior. At this point, the amplitude of the additional peak vanishes at a finite value of the tortoise coordinate. Larger values of \(\beta\) typically correspond to a wider interval of the hair parameter for which the metric is nonmonotonic.

It is important to note that the observed echo amplitude is only a few percent of the primary ringdown amplitude, consistent with the fact that the secondary peak of the effective potential is much smaller than the main peak. Although no statistically significant detections of echoes have been reported in LIGO/Virgo events such as GW150914 and GW170817~\cite{Abedi:2017isz,Westerweck:2017hus,Maggio:2019zyv}, current analyses are not sensitive to echoes with amplitudes below roughly $10\%$ of the primary ringdown signal, as such signals fall beneath the noise threshold.

The ongoing observing runs are expected to achieve improved strain sensitivity by a factor of two to three, but this improvement still appears insufficient to detect echoes arising from Proca hair. Coherent stacking of multiple events may increase the overall signal-to-noise ratio, particularly if the echoes share similar spectral features. In parallel, new Bayesian templates \cite{Tsang:2018uie} and machine-learning classifiers are being developed to support more model-independent echo searches. Unfortunately, such improvements are unlikely to be relevant for the scenario considered here, where echoes occur only within a rather narrow region of parameter space.

Future detectors offer more promising prospects. LISA’s long-duration, high signal-to-noise ratio observations \cite{Katz:2018dgn} will enable precise template matching of late-time waveform residuals, providing unprecedented opportunities to probe near-horizon physics and to distinguish black holes from exotic compact objects \cite{LISA:2024hlh}. Next-generation ground-based observatories, including the Einstein Telescope and Cosmic Explorer, will further improve strain sensitivities by roughly an order of magnitude and measure ringdown signals with significantly higher signal-to-noise ratios \cite{ET:2019dnz}. These instruments will permit population studies of potential echo candidates and may constrain relative echo amplitudes down to the \( 10^{-3}\text{--}10^{-4} \) level, making it possible to observe echoes at the amplitudes obtained in this work. A more rigorous assessment of the observational constraints, however, requires analyzing the full gravitational sector of perturbations with realistic initial conditions, which is beyond the scope of the present paper.

\section{Quasinormal modes}\label{sec:qnms}

\begin{figure*}
\resizebox{\linewidth}{!}{\includegraphics{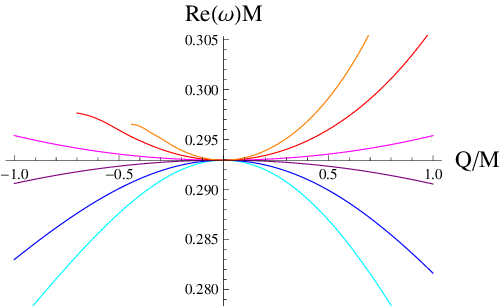}\includegraphics{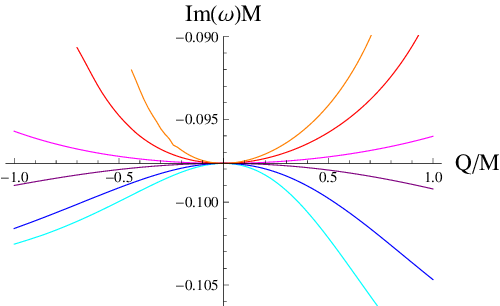}}
\caption{Real and imaginary parts for the fundamental frequency of the test scalar field \(\ell=1\) as a function of the hair parameter \(Q\) of the black hole in the vector-tensor theory (\(\beta=0\)). From bottom to top: \(\alpha=-M^2\) (cyan), \(\alpha=-0.5M^2\) (blue), \(\alpha=-0.1M^2\) (purple), \(\alpha=0.1M^2\) (magenta), \(\alpha=0.5M^2\) (red), \(\alpha=M^2\) (orange). \(Q=0\) corresponds to the Schwarzschild black hole.}\label{fig:qnmsl1b0}
\end{figure*}

\begin{figure*}
\resizebox{\linewidth}{!}{\includegraphics{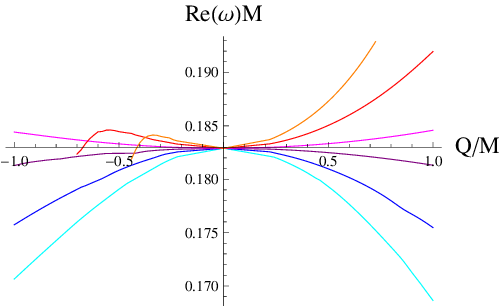}\includegraphics{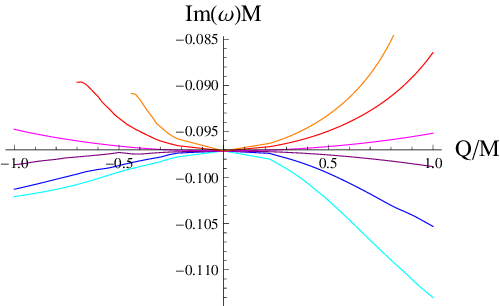}}
\caption{Real and imaginary parts for the fundamental frequency of the Dirac field \(\kappa=1\) as a function of the hair parameter \(Q\) of the black hole in the vector-tensor theory (\(\beta=0\)). From bottom to top: \(\alpha=-M^2\) (cyan), \(\alpha=-0.5M^2\) (blue), \(\alpha=-0.1M^2\) (purple), \(\alpha=0.1M^2\) (magenta), \(\alpha=0.5M^2\) (red), \(\alpha=M^2\) (orange). \(Q=0\) corresponds to the Schwarzschild black hole.}\label{fig:qnmsk1b0}
\end{figure*}

\begin{figure*}
\resizebox{\linewidth}{!}{\includegraphics{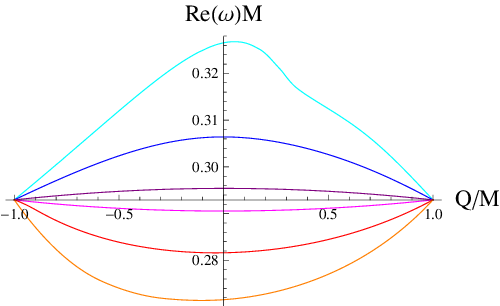}\includegraphics{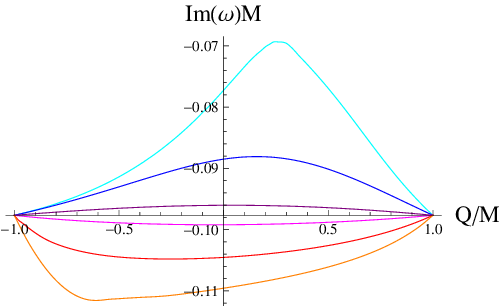}}
\caption{Real and imaginary parts for the fundamental frequency of the test scalar field \(\ell=1\) as a function of the hair parameter \(Q\) of the black hole in the theory with \(\beta=-\alpha\). From top to bottom: \(\alpha=-M^2\) (cyan), \(\alpha=-0.5M^2\) (blue), \(\alpha=-0.1M^2\) (purple), \(\alpha=0.1M^2\) (magenta), \(\alpha=0.5M^2\) (red), \(\alpha=M^2\) (orange). \(Q=\pm M\) corresponds to the Schwarzschild black hole.}\label{fig:qnmsl1bma}
\end{figure*}

In this section we analyze how the presence of black-hole hair influences the fundamental quasinormal modes.

In the pure vector-tensor theory \(\beta=0\) (see Fig.~\ref{fig:qnmsl1b0}), the effect of the hair parameter \(Q\) is straightforward. For positive \(\alpha\), both the oscillation frequency (the real part of the quasinormal mode) and the damping rate (the imaginary part) increase. For negative \(\alpha\), both decrease. The fundamental mode of the Dirac field follows the same qualitative trend (see Fig.~\ref{fig:qnmsk1b0}). However, for large negative values of the hair parameter \(Q\) and \(\alpha>0\), the real part of \(\omega\) shows a nonmonotonic dependence and does not change in such a simple one-way fashion.

When the coupling to the scalar field \(\beta\) is nonzero, the overall picture remains similar. A positive value of \(\beta\) tends to increase both the real and imaginary parts of the quasinormal frequency, while a negative \(\beta\) decreases them. A particularly interesting case when \(\beta=-\alpha\) is shown in Fig.~\ref{fig:qnmsl1bma}. Here, negative \(\alpha\) (which corresponds to positive \(\beta\)) makes both oscillation frequency and damping rate larger, whereas positive \(\alpha\) (and thus negative \(\beta\)) has the opposite effect relatively to the Schwarzschild limit (\(Q=\pm M\)).

\begin{table}
\begin{tabular}{p{0.15\linewidth}p{0.15\linewidth}p{0.15\linewidth}c}
\(\alpha\) & \(\beta\) & \(Q\) & 13th order WKB \\
\hline
 $-0.5$ & $-0.5$ & $-1  $ & $0.099-0.113 i$ \\
 $-0.5$ & $-0.5$ & $-0.5$ & $0.102-0.113 i$ \\
 $-0.5$ & $-0.5$ & $~0  $ & $0.103-0.112 i$ \\
 $-0.5$ & $-0.5$ & $~0.5$ & $0.100-0.114 i$ \\
 $-0.5$ & $-0.5$ & $~1  $ & $0.095-0.117 i$ \\
 $-0.5$ & $~0  $ & $-1  $ & $0.105-0.109 i$ \\
 $-0.5$ & $~0  $ & $-0.5$ & $0.110-0.105 i$ \\
 $-0.5$ & $~0  $ & $~0  $ & $0.111-0.105 i$ \\
 $-0.5$ & $~0  $ & $~0.5$ & $0.108-0.107 i$ \\
 $-0.5$ & $~0  $ & $~1  $ & $0.103-0.112 i$ \\
 $-0.5$ & $~0.5$ & $-1  $ & $0.111-0.105 i$ \\
 $-0.5$ & $~0.5$ & $-0.5$ & $0.116-0.099 i$ \\
 $-0.5$ & $~0.5$ & $~0  $ & $0.118-0.094 i$ \\
 $-0.5$ & $~0.5$ & $~0.5$ & $0.115-0.095 i$ \\
 $-0.5$ & $~0.5$ & $~1  $ & $0.111-0.105 i$ \\
 \hline
 $~0.5$ & $-0.5$ & $-1  $ & $0.111-0.105 i$ \\
 $~0.5$ & $-0.5$ & $-0.5$ & $0.104-0.113 i$ \\
 $~0.5$ & $-0.5$ & $~0  $ & $0.103-0.111 i$ \\
 $~0.5$ & $-0.5$ & $~0.5$ & $0.105-0.111 i$ \\
 $~0.5$ & $-0.5$ & $~1  $ & $0.111-0.105 i$ \\
 $~0.5$ & $~0  $ & $-0.5$ & $0.113-0.100 i$ \\
 $~0.5$ & $~0  $ & $~0  $ & $0.111-0.105 i$ \\
 $~0.5$ & $~0  $ & $~0.5$ & $0.113-0.102 i$ \\
 $~0.5$ & $~0  $ & $~1  $ & $0.117-0.093 i$ \\
 $~0.5$ & $~0.5$ & $~0  $ & $0.116-0.092 i$ \\
 $~0.5$ & $~0.5$ & $~0.5$ & $0.118-0.091 i$ \\
 $~0.5$ & $~0.5$ & $~1  $ & $0.114-0.081 i$ \\
 \hline
\end{tabular}
\caption{Fundamental quasinormal mode of the test scalar field (\(\ell=0\)) for various \(\alpha\), \(\beta\), and \(Q\) (\(M=1\)).}\label{tabl:qnmsl0}
\end{table}

Spherically symmetric perturbations of a test scalar field (\(\ell=0\)) show qualitatively the same tendencies. However, the dependence on the hair parameter is much weaker in this case. The parameter \(Q\) must be taken sufficiently large before the effect becomes significant beyond the error of the WKB approximation. Therefore, we provide only numerical values of the modes for large \(Q\) (see Table~\ref{tabl:qnmsl0}).

\section{Conclusions}\label{sec:conclusions}

In this work we have studied the quasinormal spectrum and time-domain profiles of scalar and Dirac fields in the background of black holes with Proca-Gauss-Bonnet hair. Our main result is that the presence of primary hair alone is sufficient to generate late-time echoes in the ringdown signal. This is in sharp contrast to the majority of previously studied scenarios, where echoes arise only after introducing additional potential barriers externally, such as from matter distributions near the horizon, clouds of quantum fields, or exotic compact objects like wormholes. Here we have shown that the intrinsic modification of the geometry by the primary hair naturally produces a secondary peak in the effective potential, which in turn leads to strong and persistent echoes.

We employed both the higher-order WKB method with Padé resummation and time-domain integration. A noteworthy technical point is that, although the WKB expansion converges asymptotically and is not necessarily improve the approximation order by order, in our case the accuracy of the Padé-resummed WKB results continues to improve when going beyond the sixth order, with the best agreement reached around twelfth to thirteenth order. This behavior departs from the common expectation that the sixth order is optimal (see~\cite{Bolokhov:2023bwm,Lutfuoglu:2025ljm,Lutfuoglu:2025hwh,Lutfuoglu:2025hjy,Konoplya:2023ahd,Dubinsky:2024aeu,Dubinsky:2024hmn,Dubinsky:2025fwv,Bolokhov:2025lnt,Bolokhov:2024ixe,Malik:2024nhy,Malik:2024qsz,Konoplya:2024lch,Skvortsova:2024atk,Skvortsova:2023zmj,Skvortsova:2024wly} for recent examples).

Our work could be extended to non-minimally coupled fields, such as gravitational and Proca perturbations. While we expect the qualitative phenomenon of echoes to persist in those cases, the quantitative characteristics will certainly differ. Another promising direction of research concerning the peculiar double-peak structure involves the study of optical phenomena and matter accretion. The first step in this direction was taken in \cite{Konoplya:2025bte}, where particle motion and the resulting black-hole shadows were analyzed for such configurations.

\bibliography{bibliography}

\end{document}